\documentclass[rnote]{aa}  
\usepackage{graphicx}
\usepackage{txfonts}
\begin{document}
   \title{Dissecting the Cosmic Infrared Background with 3D Instruments}

   \author{D.L. Clements
          \inst{1}
          K.G. Isaak\inst{2}, S.C. Madden\inst{3}, C. Pearson\inst{4}
          }

   \offprints{D.L. Clements}

   \institute{Astrophysics Group, Blackett Laboratory, Imperial College, Prince Consort Road, London SW7 2BW, UK \\
              \email{d.clements@imperial.ac.uk}     
              \and    
             School of Physics and Astronomy, Cardiff University, The Parade, Cardiff CF24 3AA, UK\\
             \email{kate.isaak@astro.cf.ac.uk}
             \and
             Service d'Astrophysique, Bat 609 Orme des Merisers, CEA Saclay, 
            91191 Gif-sur-Yvette, France\\
             \email{madden@cea.fr}
             \and
             Institute of Space and Astronautical Science, Japan Aerospace Exploration
Agency, Yoshinodai 3-1-1, Sagamihara, Kanagawa 229-8510, Japan, \& ISO Data Centre, European Space Agency, Villafranca del Castillo, PO Box
50727, 28080 Madrid, Spain\\
             \email{cpp@ir.isas.jaxa.jp}
                         }

   \date{Received ; accepted }

% \abstract{}{}{}{}{} 
% 5 {} token are mandatory
 
  \abstract % context heading (optional) % {} leave it empty if necessary 
  {The cosmic infrared background (CIB) consists of
  emission from distant, dusty, star-forming galaxies. Energetically, the CIB is very important as it contains as
  much energy as the extragalactic optical background. The nature and
  evolutionary status of the objects making up the background are,
  however, unclear.} 
  { The CIB peaks at
  $\sim$150$\mu m$, and as such is most effectively studied from space. The
  limited apertures of space-borne telescopes set the angular
  resolution that can be attained, and so even Herschel, with its 3.5m
  diameter, will be confusion-limited at this wavelengths at
  $\sim$5mJy.  The bulk of the
  galaxies contributing to the CIB are fainter than this, so it is difficult to study them without
  interferometry. Here we
  present the results of a preliminary study of an alternative way of probing fainter than the continuum confusion limit using far-IR imaging spectroscopy. An instrument capable of such observations is being planned for SPICA - a proposed Japanese mission with an aperture equivalent
  to that of Herschel and more than 2
  orders of magnitude more sensitive.}
  % Methods heading (mandatory)
  { In this paper we investigate the
  potential of imaging spectrometers to break the continuum confusion
  limit. We have simulated the capabilities of a
  spectrometer with modest field of view (2'x2'), moderate spectral
  resolution (R$\sim$1-2000) and high sensitivity. }
%  ($\sim$ a few mJy 1$\sigma$  1 hour per resolution element).}
{We
  find that such an instrument is capable of not only detecting line
  emission from sources with continuum fluxes substantially below the
  confusion limit, but also of determining their redshifts and, where
  multiple lines are detected, some emission line
  diagnostics. } 
  {3-D imaging spectrometers
  on cooled far-IR space telescopes will 
  be powerful new tools for extragalactic far-IR
  astronomy.}

\authorrunning{D.L. Clements et al.}
   \maketitle
%
%________________________________________________________________

\section{Introduction}

The discovery of the Cosmic Infrared Background (CIB) (Puget et al.,
1996; Fixsen et al, 1998) provided a key step in our understanding of the
history of the universe. The energy in the
CIB is comparable to that in the integrated UV/optical/near-IR (UVOIR) background. Since the CIB comes from obscured systems and the UVOIR from unobscured systems this means that roughly 50\% of
energy produced in the history of the Universe comes from
obscured systems. The nature of the objects making up the
CIB, however, remains unclear. Source populations that contribute
significantly to the CIB have been found in submillimetre surveys
(Blain et al., 2002, and references therein), in the far-IR (Puget et
al., 1999), and in the mid-IR (Dole et al., 2006), with as much as
70\% (Dole et al., 2006) of the CIB now believed to have been resolved into
individual sources. Much of this work, however, has been done at
wavelengths far away from the 150$\mu$m peak of the CIB, and is
reliant on extrapolations.

The Herschel far-IR mission (Pilbratt, 2003), to be launched in
2008, will be capable of deep observations around the peak of the CIB, and will be sensitive enough to resolve
a significant fraction of CIB sources at wavelengths where
their spectral energy distributions (SEDs) peak (Dole et al.,
2004). With a 3.5m-diameter mirror, Herschel will have the largest
satellite primary to date, but even this will only give an 
angular resolution of $\sim$10" at 150$\mu$m. 
The source density of CIB sources on the sky is such that confusion,
rather than instrument sensitivity, is the main limitation on the
faintness at which an individual CIB source can be detected - for
exmple, the confusion limit of the PACS instrument on Herschel
(Poglitsch et al., 2005) has been estimated to be $\sim$5mJy at the
$\sim$ 150$\mu$m peak of the CIB (Jeong et al., 2006).  Extrapolations
from the deep SPITZER surveys suggest that Herschel will therefore
only be able to resolve $\sim$50\% of the CIB (Dole et al., 2004) at
best, even at its shortest operating wavelength.

There are of course ways in which one can probe below the confusion limit by using observations at other wavelengths or through statistical approaches. This includes extraction of long wavelength fluxes at the positions of sources detected at shorter wavelengths which has proved successful in some studies (eg. Dole et al., 2006). There remains the worry that such studies can be biased against redder sources, which may be missing from the short wavelength catalogs but still contribute strongly at long wavelengths. Such issues can only be avoided by detecting sources at the long, CIB, wavelengths themselves which will require observations below the continuum confusion. In this
research note we put forward the results of work that has been done to
use an additional observational dimension to break the confusion limit - that of 
spectroscopy. 
We will demonstrate that through spectroscopy one can identify sources
that lie significantly below the traditional (contiuum) confusion
limit, and thus to probe a population that contributes significantly to
the CIB, that cannot presently be studied in any other way.
 
\section{Modeling Blank Field Far-IR Line Surveys}

\subsection{A Far-IR Imaging Spectrometer}
The key feature of an instrument designed to conduct blank field
searches for emission lines is that it must simultaneously cover a
significant field of view over a significant
spectral range with sufficient sensitivity to detect the
lines against the continuum. Imaging spectroscopy at far-infrared,
infrared and optical wavelengths can be achieved using a number of
different instrumental techniques, including imaging fourier
transform spectroscopy - as employed in the Herschel SPIRE instrument
(Griffin et al., 2004), integral field spectroscopy using a
grating and image slicer - as employed in the Herschel
PACS instrument (Poglitsch et al., 2005), using imaging 
Fabry-Perots - as employed in SPIFI (Stacey et al. 2002), and at the 
longer wavelengths with heterodyne arrays using ultra-sensitive mixers - as
employed in HARP-B (Smith et al., 2003). Each technique is more or 
less appropriate to a particular application and a particular wavelength
range/instantaneous bandwidth requirement. 

Instrument concepts based on the first two of these techniques are
currently under consideration for an imaging spectrometer (ESI, the European SPICA instrument) for the
35-210$\mu$m waveband for SPICA (Swinyard et al., 2006). For the purpose of this paper
our considerations of the instrument itself are restricted to its
deliverables, namely a data cube with a particular two-dimensional
spatial extent and angular resolution and a given spectral range and
resolution in the third (spectral) direction, with a specified
sensitivity in each spectral channel. We have based our simulations on
what we believe to be feasible for a cooled (~4K), 3.5m-diameter
primary, telescope such as SPICA: a spatial resolution of 8" as set by
the diffraction limit at $\sim$120$\mu$m; an instantaneous spectral
coverage of 60 to 210$\mu$m and a fixed channel width of
$\Delta\lambda \approx 0.176\mu$m, resulting in a spectral resolution
ranging from 350 at 60$\mu$m to 1200 at 210$\mu$m and an instantaneous
field of view of 128"x128". These instrumental parameters are very
similar to those expected in the two longest wavelength channels of
the ESI far-infrared instrument concept.

\subsection{The Far-IR Sky}

To assess the ability of our far-IR imaging spectrometer to detect
line-emitting sources in the CIB, we must model the far-IR sky. Our
starting place is the current generation of empirical number
count models that fit all current constraints on galaxy counts from
optical to submm, including the strength of the CIB. 
The model we have chosen to adopt is the burst mode
evolution model of Pearson (2006) since this provides all the necessary features we require,
and, in later work, can easily be compared and contrasted with Pearson's parallel bright galaxy evolution model. The model we use
is based on an evolving, type-dependent, far-infrared luminosity
function which makes use of a set of 11 optical-to-submm template
SEDs. The template objects range from normal, Milky-Way type galaxies,
through increasingly far-IR luminous objects to extreme ultraluminous
infrared galaxies (ULIRGs) and type 1 and 2 AGNs.  To simulate a
region of the far-IR sky we begin with a simulated catalogue of
sources with far-infrared fluxes, luminosities, redshifts and
realistic SEDs taken from the galaxy evolution model. Each source in
the catalogue is parameterised by a wavelength and a flux, which are
assigned based on the SED template type.  At this point, the template
fluxes are continuum only. Far-IR emission (and absoprtion) lines are
added to these templates using line strengths derived from LWS far-IR
spectroscopy for a range of nearby galaxies (eg. Negishi et
al. (2001)), as tabulated in Table 1. The resulting templates,
including lines, are shown in Figure 1. We only include the brightest far-IR
lines, and no MIR lines, at this stage since these will be the most useful for source
detection in the far-IR.

\begin{table}
\begin{tabular}{cc}
Continuum Template&Line Template\\ \hline
Normal Cold&Maffei 2\\
Normal&Maffei 2\\
Starburst M82&M82\\
Starburst&M82\\
10$^{11} L_{\odot}$ LIG&NGC253\\
10$^{11.5} L_{\odot}$ LIG&NGC253\\
Cold LIG&NGC253\\
Hot ULIRG&Arp220\\
Cold ULIRG&Arp220\\
Sy1&No lines\\
Sy2&No lines\\
\end{tabular}
\caption{Galaxies used to provide far-IR lines for templates.}
\end{table}

To generate the data cube that will come from observing this catalog
we compile a list of sources in our field of view. The integrated flux
contributed to each 8" pixel by all the objects that lie within it is
then calculated. The template spectrum for each object contributing to
a spatial pixel is shifted in wavelength by the source redshift and then normalised to
the flux of the source at the fiducial catalog wavelength
(40$\mu$m, the wavelength at which the original catalog was normalised). The flux at a given, observed wavelength is then
interpolated from the template and added to the appropriate wavelength
channel in the data cube. This process is
repeated for each 8"x8" spatial pixel to build up the 16 x 16 spatial
cube, each pixel made up of a 3rd dimension of 852 spectral
channels. Random Gaussian noise is then added to each of these
channels at a level appropriate to the instrument sensitivity being
simulated.

At the same time as the data cube is generated, we also produce a
'truth catalog' for the field which is comprised of 120$\mu$m fluxes (to allow comparison with the sensitivity of current and future instruments such as PACS and Spittzer),
redshifts, positions and source classifications for objects in
the 128'' x 128'' field of view. This truth catalogue is also used to
calculate the conventional '20 beams per source' confusion limit
which, for this field, is 4.3mJy.

%confusion limit 2.6mJy for the bright model.

   \begin{figure}
   \centering
   \includegraphics[angle=90, width = 10cm]{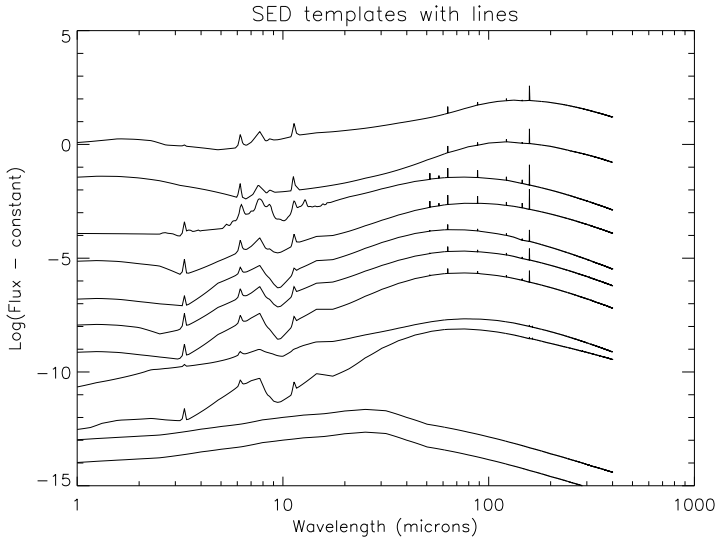}
      \caption{Spectral energy distributions used for the models}
      From the top downwards these SEDs correspond to following SED types in the models: Normal cold, Normal, starburst M82, starburst, 10$^{11}$ L$_{FIR}$ LIG, 10$^{11.5}$ L$_{FIR}$ LIG, Cold LIG, Hot ULIRG, Cold ULIRG, Sy1, Sy2. 
   \end{figure}

\section{Results of Models}

To determine the completeness of source identification via far-IR line
emission, we search the simulated data cube for sources with strong
line emission; we then compare the resulting list of sources with the
'truth catalog' derived from the original simulations. A line
detection is conducted in the following way: at each image pixel the
continuum is removed by fitting and subtracting a 4th order
polynomial; the standard deviation, $\sigma$, of the resulting
continuum-subtracted spectrum is then calculated; a source is defined
to contain a line if there exist channels in the spectrum in which the
line flux is greater than some pre-defined multiple of this rms noise,
ie. $n \sigma$. We find through comparison to the 'truth catalog' that
a reliable, ie. no false detections, list of line emitting sources is produced
when $n = 4.5$, ie. when lines with significance $\ge 4.5 \sigma$ are
selected.

Figure 2 compares the broad band image of the field at
120$\mu$m (left) with a continuum-subtracted image at a wavelength
matching the faintest detected line-emitting source, at a wavelength
of $\sim$209$\mu$m (right), for a simulation with 1$\sigma$ noise of
0.4mJy. We note that this line-emitting source has a continuum flux of
0.56mJy, nearly an order of magnitude fainter than the traditional
confusion limit. However it is still significantly detected  through its line emission. The number of sources
recovered through their line emission as a function of continuum flux,
compared to the initial distribution of fluxes in the input catalog
is shown in Fig. 3. This clearly demonstrates the power of the
technique, in that we recover all sources above the conventional
confusion limit and a substantial fraction of those with fluxes up a factor of 10  below it.
As expected, the fraction detected drops as
the source gets fainter. The potential for breaking the continuum
confusion limit through the detection of sources in their far-IR
line emission is clear.

It is possible not only to locate a source in RA-DEC via
blank-line surveys but also, very importantly, to determine its
redshift. This requires several assumptions to be made. In some
cases we detect more than one emission line in a given object, making
redshift determination quite simple. In most cases, however, only one
line is detected: in such cases the single lines detected are usually
one of two bright far-IR emission lines [CII] 157.7$\mu$m or [OI]
63.2$\mu$m. In a typical dust SED which is characteristic of dust at a
temperature of $\sim$35K, the [CII] line lies longward of the SED peak,
while [OI] lies at shorter wavelengths. If we assume that the dominant
contribution to the confused continuum emission comes from the line
emitting source, then by considering the position of the line relative
to the SED peak it is possible to guess which far-IR species is
responsible for the emission line and thus get an accurate
determination of redshift.  The efficacy of this is shown in Figure 3,
where the inferred redshifts of detected sources with $S_{120\mu
m} > 1$mJy are compared to the catalogue as a whole.
As can be seen, we recover 100\% of the redshifts
for sources at z$<$2.5 that are brighter than 1mJy ie. $\sim$5 times
fainter than the conventional confusion limit. By z$\sim$2.5, both of
the two strongest FIR lines, [CII]157.7$\mu$m and [OI]63.2$\mu$m have
moved out of the passband of our simulated instrument. We therefore do
not expect to obtain any of these higher redshifts in the current simulation.

We thus find that blind, blank-field spectral surveys are capable not only
of detecting sources well below the confusion limit {\em but of obtaining
their redshifts at the same time}. Early analysis suggests we can
distinguish between different luminosity evolution
models (the burst and bright models from Pearson (2006)) with a survey
10'x10' or less in size using these techniques. This is something not possible
using broadband number counts. Operation at higher redshifts will come from additional
narrow lines not in the current model, such as the strong [SiII]34.8$\mu$m, [SIII]33.5$\mu$m and [OIV] 24.9$\mu$m lines, some of which can be used as emission line diagnostics. Template correlation working on redshifted PAH features will also allow us to probe high redshifts.

\begin{figure}
\includegraphics[width=4.35cm]{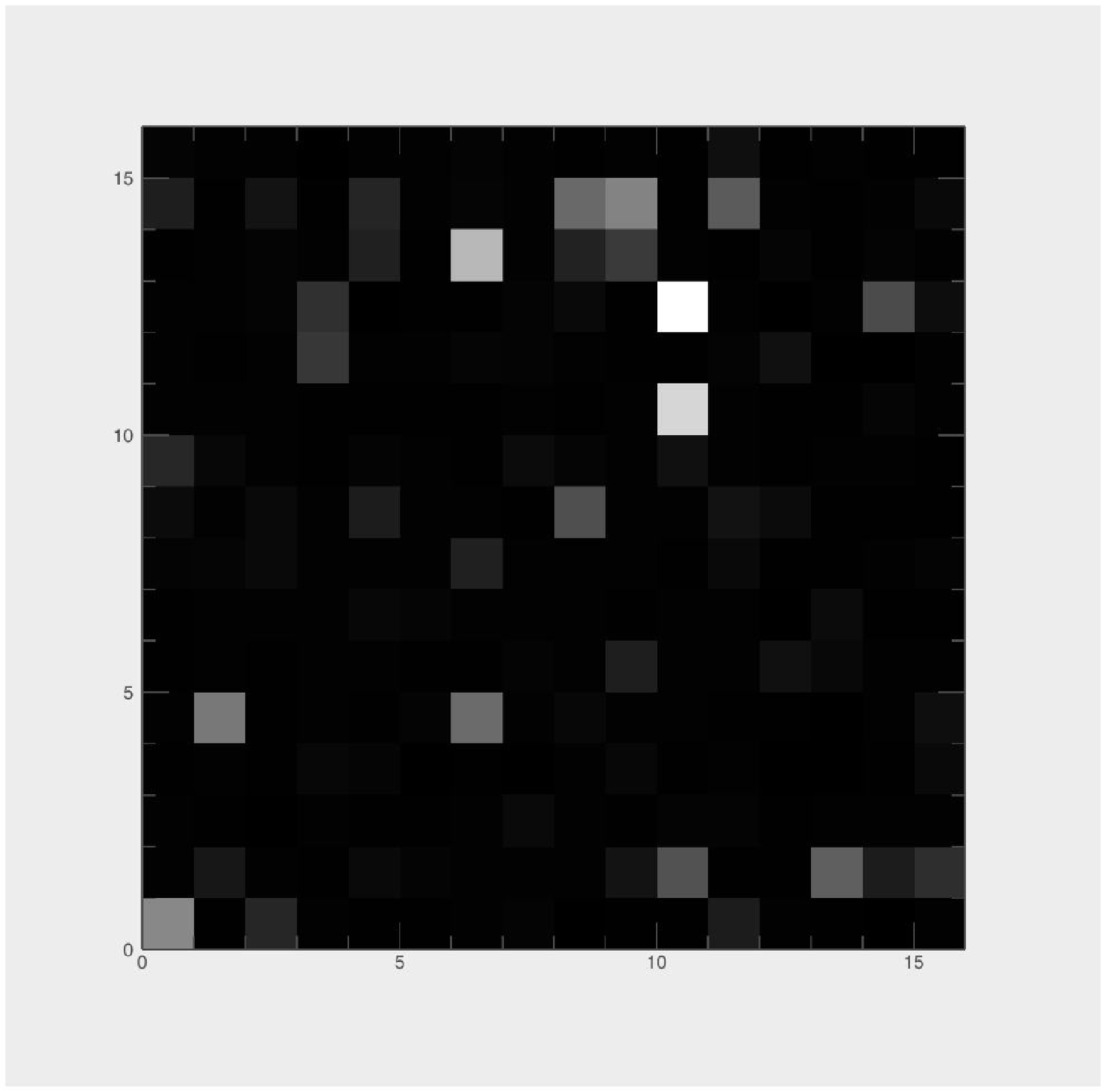}
\includegraphics[width=4.35cm]{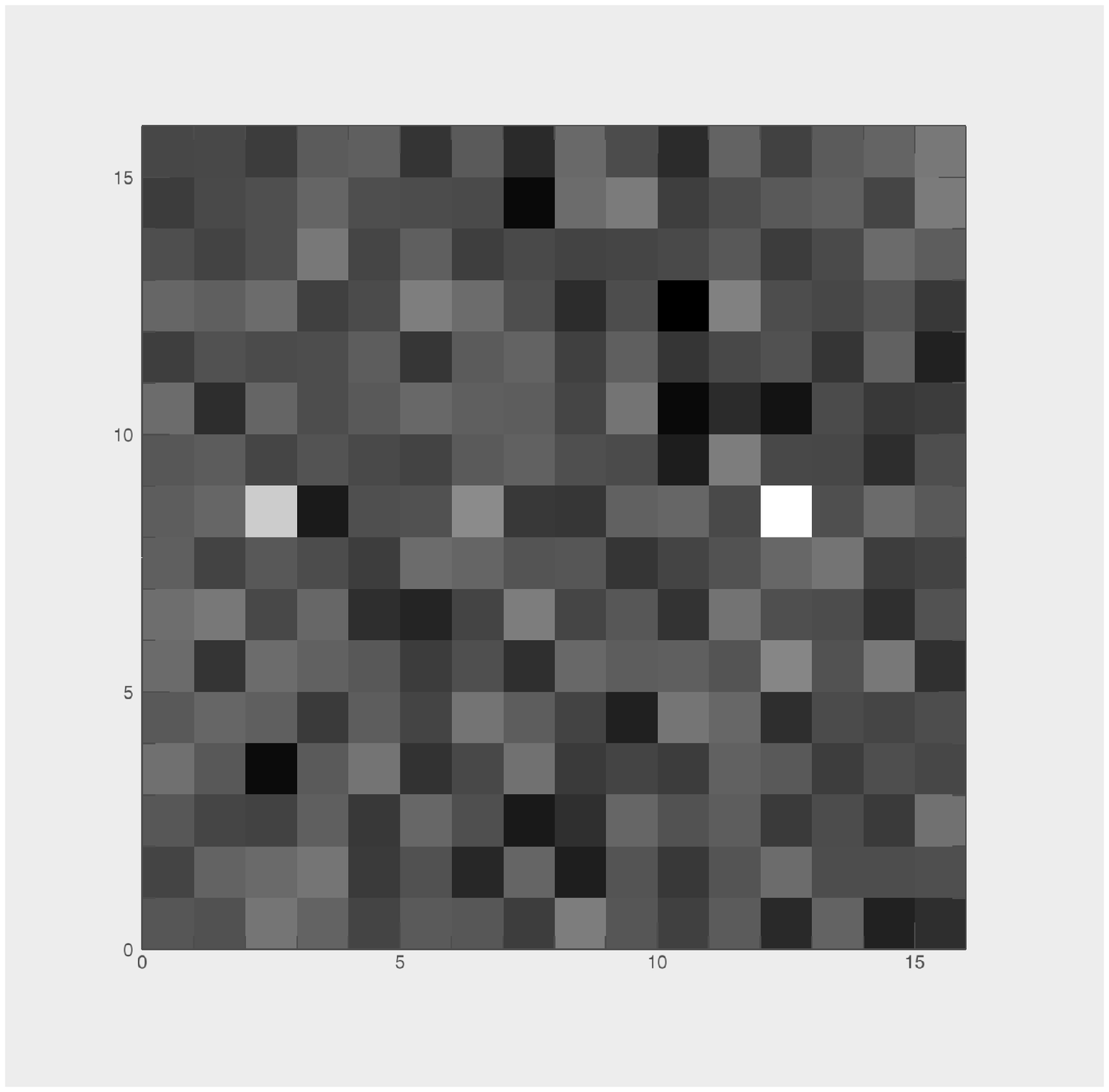}
\caption{Source detected in line below confusion limit}
{\bf Left:} Continuum 120$\mu$m image of the field. {\bf Right:} Narrow band image of the field at the wavelength at which a 560$\mu$Jy 120$\mu$m continuum flux source is detected by its line emission. 
Note that the source is clearly detected
in the line image (the rightmost bright pixel, at
location (12, 8)) while in the continuum image nothing
can be seen because of the confusion noise. Greyscales not the same, but chosen to make this effect clear. The additional source seen to the left in the line image is below our 4.5 $\sigma$ detection threshold.
\end{figure} 

  \begin{figure}
    \includegraphics[angle=90, width=8cm]{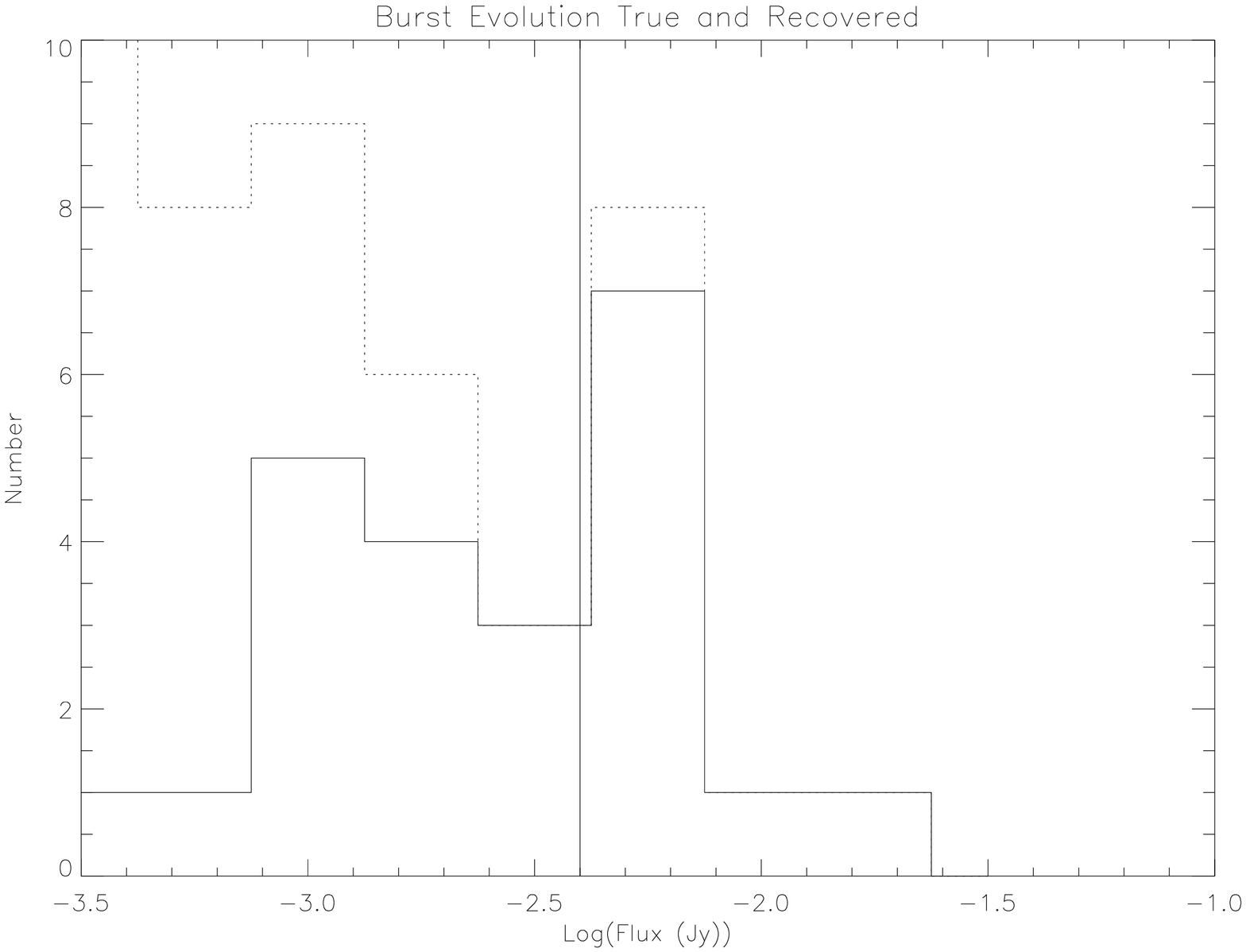}
   \includegraphics[angle=90, width=8cm]{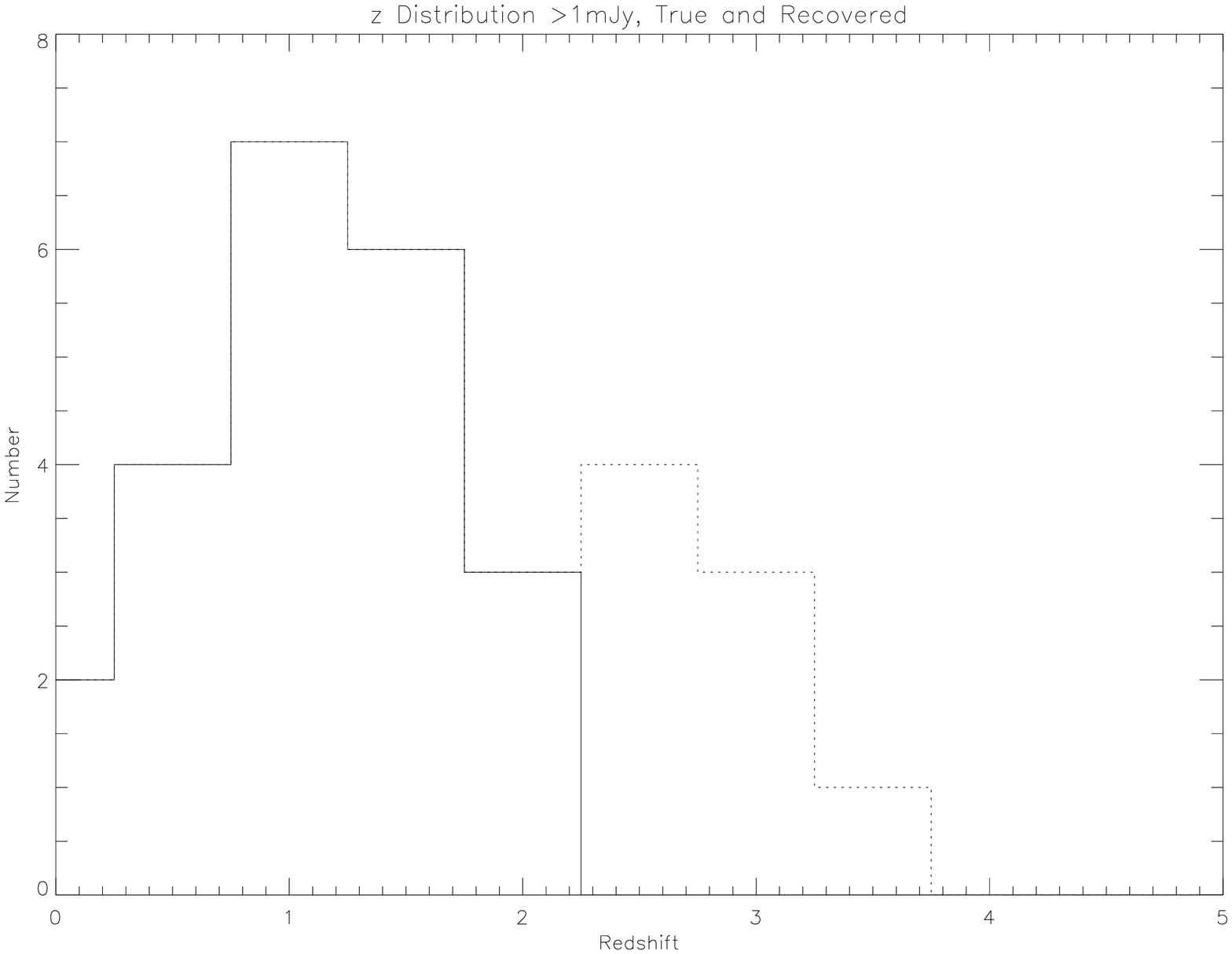}
      \caption{Recovery of Objects, with redshifts, from the FIR data cube}
      {\bf Top:} Shows the input flux distribution at 120$\mu$m from the 'truth catalog' (dotted line) and the sources detected from their line emission as a function of their 120$\mu$m continuum flux (solid line) for a 4.5$\sigma$ line detection threshold and a 1$\sigma$ noise of 0.4mJy per channel. The conventional continuum confusion limit of 20 beams per source is shown as a vertical line. Note that many sources below this confusion limit are detected, with sources having continuum fluxes nearly an order of magnitude below the limit being accessible. {\bf Bottom:} The redshift distribution from the 'truth catalog' for sources brighter than 1mJy at 120$\mu$m (dotted line) and the redshift distribution for sources brighter than 1mJy at 120$\mu$m recovered from the 4.5$\sigma$ line detected sources. Note that {\em all} sources with $z<2.5$ and $F_{120} > 1mJy$ have been recovered.
   \end{figure}

\section{Conclusions}

We have examined the potential for of imaging spectroscopic
instruments in studying the sources responsible for the CIB. The main
limiting factor in CIB studies for far-IR missions is confusion resulting from the high continuum source density
and large instrumental beams. The use of
imaging spectrometers enables the detection of sources by their line
rather than continuum emission, allowing us to detect sources whose continuum flux
is up to an order of magnitude below the conventional confusion
limit. Furthermore, a by-product of the analysis of the line and
continuum emission of sources detected in this way can be used to
determine source redshifts. We believe there is considerable mileage in deep, blank-field spectral 
line surveys in the far-infrared and will discuss the results of more extensive
modeling in a future, more substantial paper. 

\begin{acknowledgements}
This work is supported in part by PPARC, CNES, the French Programmes Nationaux and ESA, and is a product of the SPICA-ESI Phase A study. Our thanks go to all the other members of
the ESI and SPICA teams, and to the anonymous referee for many helpful comments.
\end{acknowledgements}

\end{document}